# MODELLING THE IMPACT OF ORGANIZATIONAN STRUCTURE AND WHISTLE-BLOWERS ON INTRA-ORGANIZATIONAL CORRUPTION CONTAGION

## MAZIAR NEKOVEE[1] AND JONATHAN PINTO[2]

## ABSTRACT

We complement the rich conceptual work on organizational corruption by quantitatively modeling the spread of corruption within organizations. We systematically vary four organizational culture-related parameters, i.e., organization structure, location of *bad apple*, employees' propensity to become corrupted ("corruption probability"), and number of whistle-blowers. Our simulation studies find that in organizations with flatter structures, corruption permeates the organization at a lower threshold value of corruption probability compared to those with taller structures. However, the final proportion of corrupted individuals is higher in the latter as compared to the former. Also, we find that for a 1,000-strong organization, 5% of the workforce is a critical threshold in terms of the number of whistle-blowers needed to constrain the spread of corruption, and if this number is around 25%, the corruption contagion is negligible. Implications of our results are discussed.

*Key words*: intra-organizational corruption, contagion, mathematical model, simulation study, organization structure, organizational networks, whistle-blowing, critical threshold

[1] Maziar Nekovee (m.nekovee@sussex.ac.uk) is with the School of Engineering and Informatics, University of Sussex, UK.

[2] Jonathan Pinto (j.pinto@imperial.ac.uk) is with the Business School, Imperial College, UK.

## I.     INTRODUCTION

In recent years there has been a spate of work on organizational corruption in general, and on the organizational contagion processes that result in corruption getting normalized or institutionalized in particular (Ashforth & Anand 2003; Brief, Buttram, & Dukerich, 2001; Darley 2005; Moore 2009; Pinto, Leana, & Pil, 2008). Although this work has identified a variety of mechanisms through which corruption could spread across an organization, quantitative studies of these dynamics and their consequences have been relatively neglected. In parallel, there has been a wealth of quantitative studies about diffusion in general and contagion in particular (Dodds & Watts 2005) which have investigated the spread of, *inter alia*, fads and fashions (Bikhchandani, Hirshleifer, & Welch, 1992; Hirsh 1972), unpopular norms (Bicchieri & Fukui, 1999), rumors (Nekovee, Moreno, Bianconi, & Marsili, 2006), sentiment contagion (Zhao et al. 2014), management practices (Abrahamson & Fairchild 1999), organizational forms (Lee & Pennings 2002), and civil service procedures (Tolbert & Zucker 1983), among others. These studies have been conducted across a variety of units and levels of analyses, e.g., individuals (Bikhchandani et al. 1992; Hirsh 1972), organizations (Abrahamson & Fairchild 1999), industry (Lee & Pennings 2002), cities across a State (Tolbert & Zucker 1983), and States of a country (Goel & Nelson 2007). However, intra-organizational contagion has been relatively neglected.

Our paper is at the intersection of corruption and contagion literature streams because we quantitatively model corruption contagion within organizations.  Although there has been substantial conceptual and qualitative exposition of the social contagion processes that result in organizations becoming corrupt, there has been almost no quantitative work on this subject (see Chang & Lai 2002 for an exception). For instance, Pinto et al. (2008: 688) conceptualize an *organization of corrupt individuals* as one that results when personally corrupt behaviors cross a critical threshold but they do not precisely specify this level or point at which the phenomenon escalates from the individual-level to the organization-level. They follow Andersson and Pearson (1999) who suggest that when the number of incivility spirals reach a critical threshold it may result in "uncivil" organization, again without quantifying the threshold or tipping point.

We bridge this gap by quantitatively modeling the dynamics of corruption spreading in organizations to investigate the following questions: What is the critical threshold in terms of corruption probability (i.e., employees' propensity to be corrupted) which, once crossed, will result in the entire organization being corrupted?; What are the evolution or contagion dynamics of corruption from a single *bad apple* to an organization-wide phenomenon?; What proportion of the organization should be potential whistle-blowers to prevent it from being corrupted? The model we have developed is general and is applicable to any organizational structure but as a first step, we investigate its dynamics on purely hierarchical organizational networks, considering both tall and flat structures.

The rest of this paper is organized as follows. In Section II we develop a stochastic model for corruption dynamics in organizations. In Section III we perform extensive Monte Carlo simulations of our model on a set of purely hierarchical organizational networks, and describe the results of the corruption spreading dynamics, i.e., the corruption threshold, the impact of organizational hierarchy, and the impact of whistle-blowers on the corruption spreading. We conclude this paper in Section IV with a discussion of the implications of our findings for future theorizing and practice.

## II. MODELLING CORRUPTION CONTAGION

### II.1 Phenomenology

Contagion is defined as the spreading of an entity or influence between individuals in a population via direct or indirect contact (Dodds & Watts 2005) and social contagion (Goldstone & Janssen, 2005: 427) is defined as "the spread of an entity or influence between individuals in a population via interactions between agents. Examples are the spreading of fads, rumors, and riots." In this paper we focus on one form of social contagion; the spread of a corrupt practice. Corruption is generally defined as the misuse of a position of authority for private or personal benefit (Shleifer & Vishny 1993).

One of our key objectives is to derive a point-estimate of the "critical threshold" (Andersson & Pearson 1999; Pinto et al. 2008), or "tipping point" (Gladwell 2000), which when breached results in the corrupt practices effectively pervading the entire organization. Identifying the critical threshold is important because once corruption pervades an organization, the organization will almost certainly decay and die, resulting in enormous social and economic costs. We bound our paper around the definition of organization of corrupt individuals (Pinto et al. 2008: 688), which is "an emergent, bottom-up phenomenon in which one or more mesoscale processes facilitate the contagion (and sometimes the initiation as well) of personally corrupt behaviors that cross a critical threshold such that the organization can be characterized as corrupt."

For corrupt practices to diffuse widely within an organization, the organizational culture would necessarily be complicit. We therefore include four key organizational-culture related parameters in our simulation studies. Our modeling parameters include organization structure, *bad apples*, individuals' propensity to be corrupted, and whistle-blowers.

*Organization structure:* Corruption could permeate both through proximity, and aspects like interdependent relationships and mentoring programs both vertically and horizontally. With regard to vertical corruption contagion, Bovasso (1996) found that individual adoption of attitudes and behaviors is influenced by those who have power over them, and this could result in crimes of obedience (Hamilton & Sanders 1999). With regard to horizontal corruption, Greve (1995: 450) states that "in a decentralized organization with many decision-making nodes, horizontal contact among decision makers within the organization is likely and can lead to contagion of practices within the organization." Both forms of contagion could co-exist and reinforce each other. Jones and Kavanagh (1996) found that when authority figures and peers both exert an influence towards corrupt behavior, their effect is amplified. However, it is not clear from the literature whether corrupt practices would permeate taller organizations faster, and to a greater extent, than flatter organizations, or if it would be vice versa. To attempt to address this issue we model corruption contagion with regard to both taller structures and flatter structures.

*Location of "bad apple"*: In the *bad apples* perspective on corruption, individual characteristics are assumed to be the primary force influencing unethical behavior (Brass et al. 1998; Felps et al. 2006; Gino et al. 2009; Moore 2009; Trevino & Youngblood 1990). Further, Bettenhausen and Murnighan (1985) found that when norms are not immediately apparent, individuals tend to transpose norms from past experience in similar situations. This notion resonates with management research that utilizes modeling techniques. Puranam and Swamy (2010: 4) state, "do not assume that the agents always commence from a state of agnosticism about the true state of affairs. Instead of a tabula rasa, interacting agents often bring with them their own mental models of the situation (Rouse & Morris 1986)." Thus, a new recruit who has been engaging in corrupt practices in his or her previous organization may transpose them into the current organization and thereby become a *bad apple*.

According to Greve (1995), a theoretical interest pioneered by network studies is that the location of actors in a social structure leads to heterogeneity in contagion (Burt 1987; Galaskiewicz & Burt 1991; Marsden & Friedkin 1993; Strang & Tuma 1993). For instance, centrality in the information structure is associated with greater likelihood of contagion (Coleman et al. 1966). In a branching network (i.e., organization structure) it is obvious that the higher the *bad apple* is located in the hierarchy, the greater the likelihood, speed, and pervasiveness of corruption throughout the organization. However, the differential impact of the *bad apple* being at various levels of the hierarchy is not clear, especially across organizations of differing heights/breadths. Hence, we model the time evolution of corruption that emanates from a single *bad apple* varying the hierarchical level at which it is located.

*Propensity to become corrupted (or "corruption probability"):* In order for corruption to spread in an organization the presence of *bad apples* is necessary but not sufficient. It is also required that the rest of the organization must be influenceable and adopt the corrupt practice. Following research in personality psychology, we take the view that individuals differ in their propensity to be influenced. One of the key personality constructs that has influenceability at its core is self-monitoring. Self-monitoring is the extent to which a person observes their own expressive behavior and adapts it to the demands of the situation (Gangestad & Snyder 2000). According to Pinto et al. (2008: 691), "high self-monitors have more variability in attitudes and behavior, pay more attention to others' expectations, and have lower commitment (Day, Schliecher, Unckless, & Hiller, 2002) and they are more likely to engage in unethical behavior (Ross & Robertson 2000)." We model employees' propensity to become corrupted as a probability ("corruption probability").

*Number of whistle-blowers:* Although diffusion and contagion studies usually focus on the take-up and spreading, rather than on the inhibition and dropping-off (Strang & Macy 2001), in this paper we also include an inhibitory factor, i.e., whistle-blowing. Whistle-blowing is an important antidote to corporate corruption (Boyle 1990; Paul & Townsend 1996). Near and Miceli define whistleblowing as "the disclosure by organization members (former or current) of illegal, immoral, or illegitimate organizational acts or omissions to parties who can take action to correct the wrongdoing" (1985: 4). Thus, if an employee who is a whistle-blower becomes aware that a corrupt practice is being engaged in, he or she would blow the whistle and the contagion would be arrested. We do not specify whether the whistle is blown anonymously or not (Nayir & Herzig, 2011).

II.2 Mathematical Model

We follow the standard assumption in all mathematical models of contagion that the infection probability is independent and identical across successive contacts (Dodds & Watts 2005). Thus, we have developed an *independent interaction Poisson model* rather than a *threshold model*, which asserts that an individual can only become infected when a certain critical number of exposures has been exceeded, at which point infection becomes highly probable (Dodds & Watts 2005). Our modeling methodology has two components: (1) a model for the underlying organizational contact network along which the contagion spreads; and, (2) a mathematical formulation of the phenomenology of corruption spreading. In the following each of these components are described.

The model described below is adapted from the rumor-spreading work by Nekovee et al. (2006) and is an attempt to formalize and simplify the behavioral mechanisms in terms of a set of simple but plausible rules. In formal rumor-spreading models (Daley & Kendal 1965; Maki 1973; Nekovee et al. 2006), a closed population is subdivided into three groups; those who are ignorant of the rumor ("ignorants"), those who have heard it and actively spread it ("spreaders"), and those who have heard the rumor but have ceased to spread it ("stiflers"). Early work on modeling rumor spreading (Daley & Kendal 1965; Maki 1973) assumed a homogeneously mixed population. This approximation is appropriate when the population is relatively small such that one can assume that each individual can directly contact every other individual in the population, i.e. the underlying contact network is well-approximated by a fully connected network where every individual can contact every other individual in the population. Nekovee et al. (2006) extend the model to the case where contacts can take place along the links of a social network with an arbitrary connectivity structure. We adapt the model by Nekovee et al (2006) in the following manner, i.e., the spreading of corruption in an organizationally bounded population of individuals.

## Model of corruption spreading

We consider a population of individuals within an organization, e.g. a firm or corporation. The population is subdivided into two groups; those who are ignorant of the corrupt practice ("innocents"), and those who are engaging in the corrupt behavior ("corruptors") and attempt to spread it. The corrupt behavior permeates through the organization by *directed* contacts of the corrupted with the innocents. We assume that the contacts take place via the organizational contact network. We also assume that there are already some corrupt individuals, following previous work (e.g., Blanchard, Krueger, & Krueger, 2005) showing that corruption rarely emerges out of nothing, but it is usually related to some already corrupt individuals or environment which may "infect" the susceptibles.

In our model, the susceptibles or innocents, as we term them, may become corruption-aware by direct interaction with corruptors. On becoming corruption-aware, the innocents could respond in one of three ways: (1) get infected (i.e., they adopt the corrupt practice as well), and thus become "corruptors", (i.e., they actively attempt to spread the corruption to other innocents through direct interaction); or, (2) do not get infected, and are termed "uprights" because though no longer innocent (i.e., they are aware of the corrupt practice), they do not adopt the corrupt practice; or, (3) blow the whistle, and are termed "whistle-blowers", who on becoming corruption-aware report the wrongdoing to the authorities who effectively stop the corrupt practice. We note that whistle-blowing is a feature which is not present in the rumour spreading model of Nekovee et al (2006). Another difference is that in the model of Nekovee et al (2006) an infected node may become a "stifler" and stop spreading the rumor after being contacted by another infected node, a feature which cannot be rationalized in the context of corruption spreading.

The whistle-blowing process in reality is never this simple (Vandekerchkove & Lewis, 2011) or this effective (e.g., Near & Miceli 1996), and would depend on numerous situational factors (Robinson et al. 2011). We are making this simplifying assumption for the purposes of our modeling. We assume that upon the reporting of a corrupt individual by a whistle-blower, the corrupt individual is immediately removed from the organization and replaced by an upright member. This approach also resonates with the business ethics literature. For instance, Zyglidopoulos and Fleming (2007) also parse individuals into innocents (bystanders and participants), corrupted (active rationalizers and guilty perpetrators), and whistle-blowers.

Our mathematical formulation of the dynamics of the above model is as follows. At each timestep, members of an organization could be in one of the following four behavioral states: innocent, corruptor, upright, or whistleblower. An innocent is an individual who is unaware of the corruption practice, whereas a corruptor is an individual who practices the corrupt behavior and attempts to spread it to other members of the organization. An upright is an innocent individual which has been the subject of an unsuccessful attempt by a corruptor but has not adopted the corruption practice, a whistleblower is a node which upon being contacted by a corruptor cause the corruptor to be removed from the organization and replaced by an innocent. The contacts between the corrupted and the rest of the organization are governed by the following set of rules: (1) a corruptor contacts, i.e., attempts to influence an innocent, at each timestep ; (2) whenever a corruptor contacts an innocent, the innocent could become corrupted with probability $\lambda$; (3) if, after contacting an innocent, the innocent does not become a corruptor, the corruptor may become spontaneously innocent with probability $\delta$; (4) if a corruptor contact a whistleblower it will be removed with probability one from the organization and replaced by an innocent member due to the above-mentioned whistle-blowing process.

In the above, the first rule models the tendency of corrupt individuals to justify their behavior by influencing others to do likewise. The second rule models the tendency of individuals to adopt a corrupt practice, particularly if it is in their personal interest. The third and fourth rules indirectly model the ethical culture of an organization. If the ethical culture of the organization is strong then the probability of a corruptor infecting the innocents should reduce with every failure to infect an innocent, and the corruptor should either realize that the ethical behavior is the best course, at least while he or she works for that organization, or is removed from the organizational by the whistle-blowing process

## **Model of organizational network**

Gulati and Puranam (2009) distinguish between the formal organization (i.e., the normative social system designed by managers) and the informal organization (i.e., the emergent pattern of social interactions within organizations), both of which simultaneously co-exist and jointly affect organizational performance. In this paper we have assumed that contacts only take place along the links of the formal organization, i.e., organization structure.

Our model for the organizational network currently takes as its point of departure the simplest version of organizational networks: a pure hierarchical organizational tree with branching ratio $k$ and $L$ levels of organizational hierarchy. The number of nodes (agents) in this network is given by $N = (k^L - 1)/(k - 1)$. Furthermore, the average degree of the network is obtained from

$$\bar{k} = 2k \frac{k^{(L-1)} - 1}{k^{L-1}}$$

We note here that the above model of organizational network does not incorporate the informal social connections between members which may exist across the hierarchy. Indeed, the formal hierarchical network may be considered as a well-defined backbone on top of which an overlay social network could be superimposed using, for example, the network construction algorithms described in Dodds et al. (2003). However, unlike the organizational hierarchy which is very well defined, it is rather difficult to map out the overlay social network among the members, and different choices may result in hugely different network structures. For this reason, we will limit ourselves in the current study to investigating corruption dynamics taking place along the formal organization network. Consequently, by choosing different values for the parameters $k$ and $L$ we are able to model a range of organizational network structures, from "flatter" (corresponding to choosing small values of $L$ and large values of $k$) to "taller" (corresponding to choosing small values of $L$ and large values of $k$).

### III.   SIMULATION STUDIES

We performed Monte Carlo simulation studies of the afore-mentioned model on three synthetically generated organizational networks which were characterized by ($k$=3, $L$=7), ($k$=4, $L$=6) and ($k$=10, $L$=4). These networks were chosen such that the total number in the organization for all was roughly the same (i.e., around 1000) as well as having very similar average degree ($\bar{k} = 2$), but with the network structure becoming less hierarchical and more flat as $L$ was decreased and $k$ was increased at the same time. In each simulation we assume that corruption starts from a single corrupted individual and then spreads in the organization due to direct contacts between corruptors and innocent members. Since the dynamics of corruption are stochastic, each such corruption spreading event was repeated over 100 Monte Carlo runs in order to obtain meaningful statistical averages. Furthermore, corruption would spread differently depending on the position of the initial corruptor node in the organization, and the results were also averaged over events starting from 10 different randomly chosen initial corruptor nodes.

In addition, in order to analyze the impact of the position of the initial corruptor in the organization on the spreading, we also performed an additional set of simulations where we specified the "rank", $l$, of the initial corruptor (i.e., a node at the top of the organizational hierarchy has rank $l = 1$, while a node at the bottom of the hierarchy has rank $l = L$).

All simulations were performed until the number of corrupt individuals became stabilized around a stationary mean. At each timestep of the simulation each corrupt individual attempts to corrupt all other innocent individuals to whom it is connected via an organizational link. If the corrupted individual is successful in such an encounter, then the innocent individual becomes corrupted at that timestep and will attempt to spread the corruption from the next timestep. On the other hand, if the corrupted individual is unsuccessful in its attempt, it may become innocent with probability $\delta$ at the next timestep. Finally, if a corrupt individual attempts to corrupt a whistle-blower then it will be removed at that timestep and replaced by an innocent in the next timestep.

III.1 Critical Corruption threshold

In the first study we model corruption contagion across three organization structures that vary with regard to how tall or flat they are. We fix the value of delta equal to 1 and investigate how varying the corruption probability $\lambda$ impacts the final proportion of corrupted individuals. We note that in our model delta is the probability that a corruptor becomes innocent after an unsuccessful attempt to turn an innocent individual into a corrupt individual.[3]

---

[3] *We note that even with delta set to one there is a clear difference between this mechanism and the whistle-blowing mechanism in the model since if a corrupt individual contacts a whistle-blower it is immediately replaced by an innocent. On the other hand, it the corrupt individual contacts an innocent it will only turn into innocent if its attempt of corrupting the individual is unsuccessful.*

In Figure 1 we show how the final proportion of corrupted individuals in each organization changes as the corruption probability increases. As can been seen from Figure 1, as λ is gradually increased from zero towards one there is a transition from the state in which corrupt behavior dies out due to unsuccessful contact with innocent nodes and is unable to permeate the organization. However, above this threshold the corruption is able to spread in a significant proportion of the organization, with the final number of corrupted individuals increasing as corruption probability increases. This result indicates that our model has a *critical corruption threshold* below which corrupt behavior dies and above which it is able to spread to a significant proportion of the population. We have verified numerically that the threshold is approached with zero slope, i.e. the transition to the corrupt state happens smoothly.

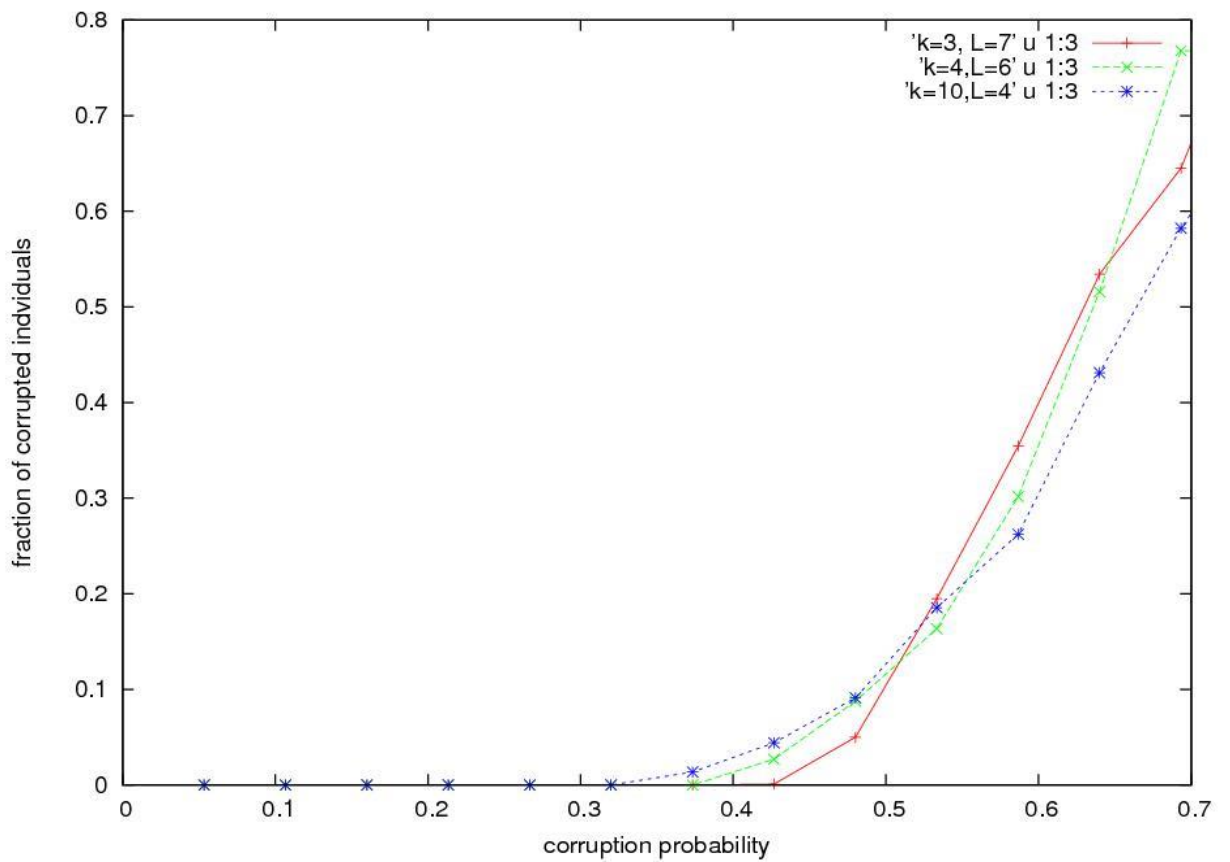

**Figure 1 The final fraction of the corrupted individuals as a function of corruption probability is shown for three organizational structures.**

We have found that the value of the critical corruption threshold depends on the topology of the underlying organizational network along which corruption spreads, i.e., the organization structure. In particular, as can be seen from Figure 1, the threshold is lowest (around $\lambda = 0.32$) for the flattest organization structure ($k$=10, $L$=4) and increases with increasing organization hierarchy to around $\lambda$ =0.38 for the moderately tall structure ($k$=4, $L$=6), and around $\lambda = 0.42$ for the tallest structure ($k$=3, $L$=7). This result suggests that organizations with a flat hierarchy, such as professional service firms may be more susceptible to organizational corruption than organizations with taller structures, such as manufacturing firms. This finding seems to resonate with reality, going by the number of investment banking firms that have been indicted on corruption charges.

However, once the corruption takes off, the final number of corrupted individuals does not seem to show a systematic dependence on network hierarchy, and this is an aspect that we can investigate in future research. Although the flattest structure has the lowest corruption contagion threshold, it spreads at a lower rate than in the other two taller structures and at the end of the simulation has resulted in the corruption of approximately 58% of the workforce, as compared to approximately 65% for the tallest structure, and approximately 78% for the moderately tall structure. This finding could be explained by the fact that flatter structures result in looser coupling (Reichman 1993) and "compartmental insulation" (Goffman 1970: 78) and this acts as a barrier to the pervasion of the corrupt practice.

That organizations with moderately tall structures can get corrupted to a greater extent than those with very flat or very tall structures is a counter-intuitive result (the intuition would be that it would be lower) which can be investigated in future research.

III.2 Impact of "*Bad Apple*" location on corruption contagion

Next, we investigate how the location of the initial corruptor member in a tall organization structure (*k*=3, *L*=7) impacts the dynamics of spreading at two different corruption probabilities, λ=0.6 in Figure 2, and λ=0.5 in Figure 3. The figures depict the time evolution of the number of corrupted individuals. Results are shown for scenarios when the initial corruptors are at different levels in the organizational hierarchy, with *l*=1 corresponding to an individual at the top and *l*=7 corresponding to an individual at the bottom of the hierarchy.

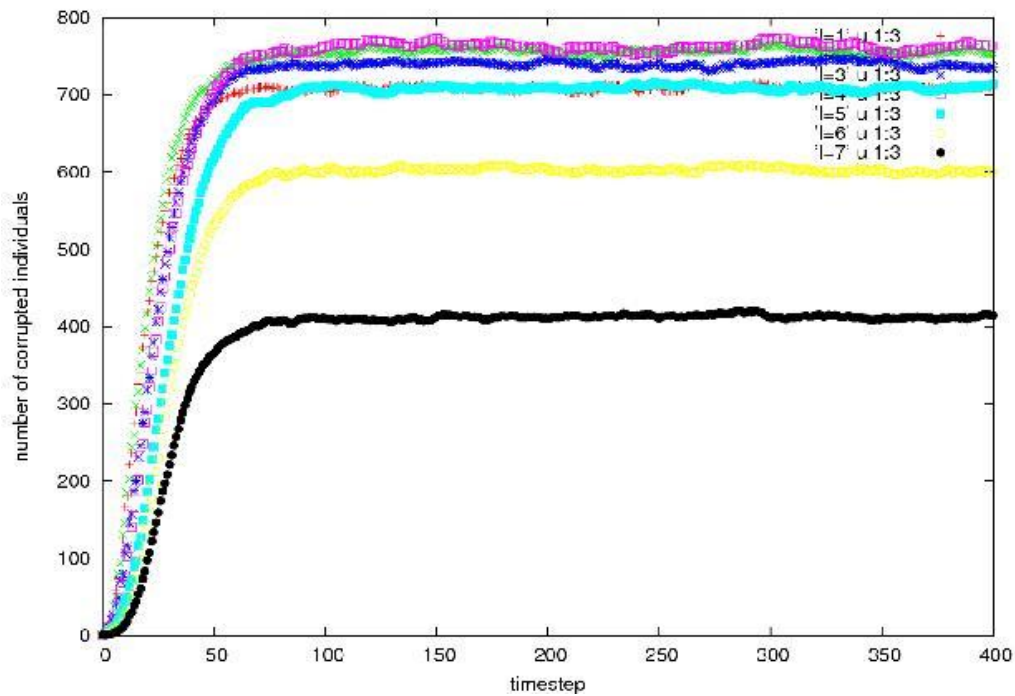

**Figure 2  Time evolution of corruption spreading is shown for the tallest structure and corruption probability = 0.6. Results are shown for corruption starting from a *bad apple* located at different position in the organizational hierarchy, varying from *l* = 1 (top of organization) to *l* = 7 (bottom of the organization).**

In Figure 2, the value of λ is set at 0.6, which is much higher than the corruption probability threshold for this type of structure (i.e., around 0.42, from Figure 1). It seems that regardless of the location of the initial corruptor in the hierarchy, the corruption spreads rapidly through the organization and results in a large proportion of the organization's workforce getting corrupted. It can be seen from Figure 2 that the position of the initial corruptor in the organizational hierarchy can greatly impact the spreading of corruption. Specifically, we see that, as expected, corruption grows slower and "infects" a much smaller proportion of the organization when it starts from the bottom. However, for this particular value of λ, it appears that the differences between spread of corruption when the initial corruptor is at various levels of the hierarchy (apart from the lowest levels, l=6,7), is not very marked. We note that due to the hierarchical structure of our networks the majority of the nodes belong to the lowest levels, e.g. in the above network 727 nodes out of a total of N=1039 nodes are located at l=7. Therefore, the final fraction of corrupted individuals as shown in Figure 1, which is obtained by averaging over results obtained by starting the corruption from 10 randomly chosen individuals in the networks, is always dominated by the results for l=7 and l=6. Thus, the results in Figure 2 are consistent with the corrected Figure 1. This observation also holds for the results shown below in Figures 3 and 4.

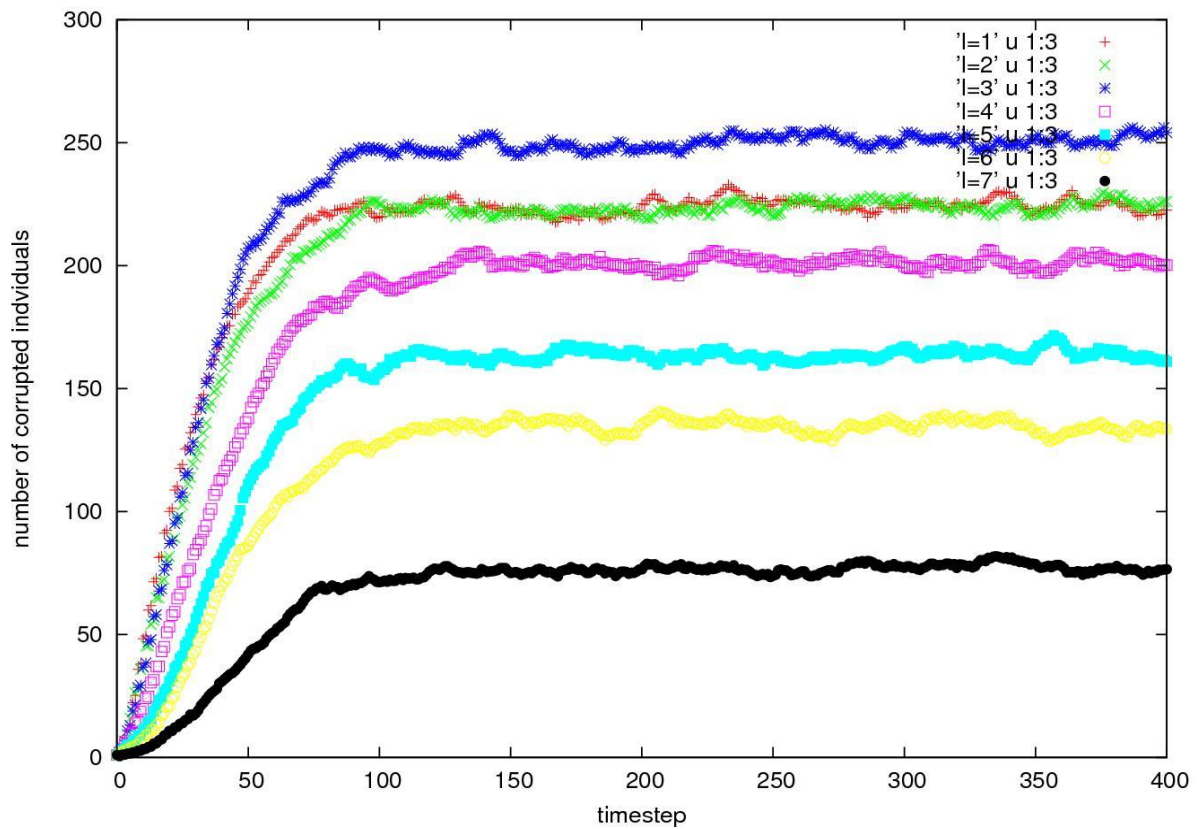

**Figure 3** Time evolution of corruption spreading is shown for the tallest structure and corruption probability = 0.5. Results are shown for corruption starting from a *bad apple* located at different position in the organizational hierarchy, varying to *l* = 1 (top of organization) to *l* = 7 (bottom of the organization).

In Figure 3, the value of λ is set at 0.5, i.e., closer to the corruption threshold (0.42) and the impact of the position of the initial corruptor on the dynamics is much more pronounced than that for λ=0.6 scenario. Firstly, as would be expected, a much smaller proportion of the organization's workforce is eventually corrupted in all seven conditions. However, despite this lower range compared to the λ=0.6 scenario, the differences between the seven conditions are much more marked. Taken together, Figures 2 and 3, suggest that in tall organization structures, a *bad apple* at any hierarchical level other than the bottom-most will result in widespread corruption contagion regardless of the corruption probability of individual employees. Further, a reduction of 0.10 in the corruption probability results in a halving of the overall pervasion of the corruption contagion. This implies that in taller structures, e.g. manufacturing firms, recruitment and selection should not only focus strongly on hiring high-integrity individuals, but also take special care in this regard when hiring employees at levels other than the lowest level.

In order to compare the differential impact of hierarchy, we conduct a similar simulation study to Figure 2 but for the flattest rather than the tallest organization structure. In Figure 4, results are shown for the flattest structure (i.e., *k*=10, *L*=4) with λ=0.6.

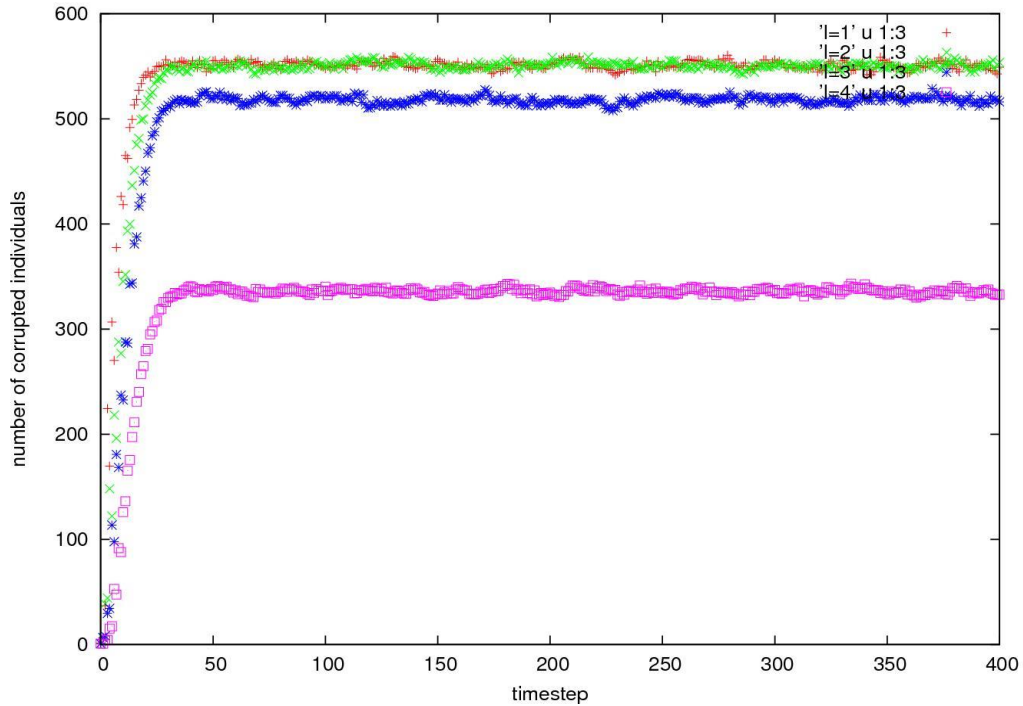

**Figure 4 Time evolution of corruption contagion and for different choices of the position of the initial corrupted member (*bad apple*) in organizational hierarchy is shown for the flatter structure and corruption probability fixed at λ=0.6.**

Once again, it can be seen that moving the initial corruptor from the bottom of the hierarchy to higher levels results in a very pronounced increase in both the speed of spreading and the final number of corrupted nodes. However, there is marked difference between the spread of corruption when the initial corruptor is at the lowest hierarchical level as compared to the higher three levels.

Comparing the spread of corruption for λ=0.6 between the tallest structure (Figure 2), and the flattest structure (Figure 4) we can see that in the latter case, the corruption does not permeate the organization to the same extent as in the former case. This reinforces our conclusion from Figure 1 that corruption spreads more slowly in flatter structures as compared to taller structures.

III.3 Impact of Whistle-blowers on Corruption Contagion

The impact of whistle-blowers on preventing widespread corruption in organizations is investigated via simulations performed for the network corresponding to (*k*=4, *L*=6), i.e., the moderately tall organization structure. For this study we randomly designate a proportion *p* of the individuals as whistle-blower and then simulate the resulting corruption dynamics following the same Monte Carlo method as described before. Figure 5 shows how increasing the proportion of whistle-blowers, at levels of 1%, 5%, 10% and 20% of the organization's workforce, impacts the final proportion of corrupted individuals.

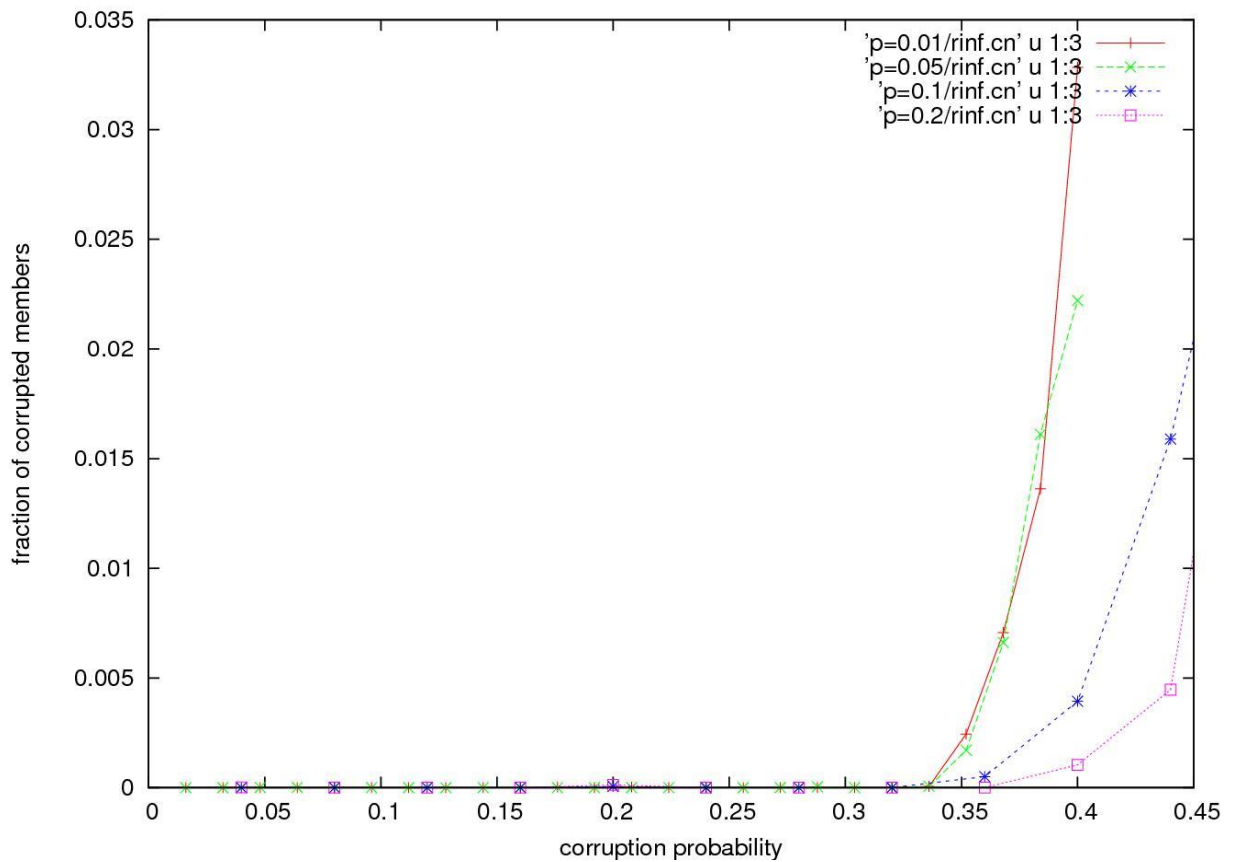

**Figure 5 Impact the fraction of whistle-blower on onset of corruption contagion threshold is shown for the moderately tall network (corresponding to k=4, L=6).**

It can be seen that if the proportion of whistle-blowers is less than 5% then the impact of whistle-blowing is insignificant. However, once the number of whistle-blowers is increased above 5% the presence of such individuals could prevent widespread corruption altogether by shifting the value of the corruption threshold upwards.

In Figure 6 results are shown for the final proportion of corrupted individuals as a function of *p*, and with the corruption probability fixed at λ=0.6.

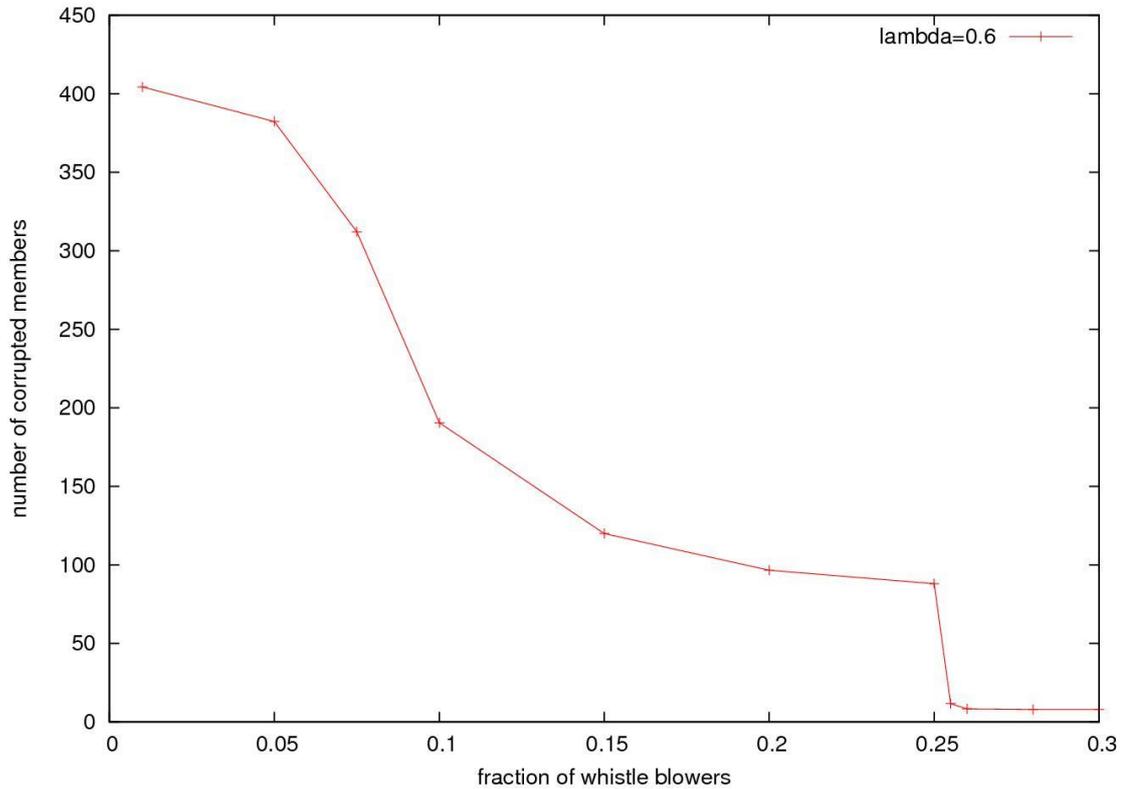

**Figure 6: Impact of the fraction of whistle-blowers on the final number of corrupted individuals is shown. for the moderately tall network (corresponding to k=4, L=6).**

Increasing the proportion of whistle-blowers initially results in a gradual decrease in the final size of corruption. However, it can be seen that there is phase transition (tipping) at around *p*=0.25, i.e., when a quarter of the organization members are whistle-blowers, above which we see a dramatic decrease in the final size of corruption. This is an interesting result as it indicates that there is a critical value for the proportion of whistle-blowers in the organization above which the action of such individuals can prevent widespread corruption.

III.4 Implications for practice.

Firstly, it seems that when corruption probability is high, then regardless of the type of organization structure, the corruption will permeate the entire organization. This underscores the importance of testing prior to selection. In terms of selection testing, researchers have found that unethical behaviors can be predicted both from specific integrity or honesty tests (Bernardin & Cooke 1993; Ones, Viswesvaran, & Schmidt, 1993) and from general personality instruments (Lee, Ashton, & de Vries, 2005). This is even more important for organizations with flatter structures, such as professional service organizations (e.g., lawyers, accountants, consultants, investment bankers), because the corruption probability threshold for these organizations is lower than it is for organizations with taller structures. Unfortunately, real-life evidence suggests that professional service organizations may not be cognizant of this risk, and their focus on business objectives rather than on their professional code of conduct could increase their likelihood of being corrupted to above the critical threshold, with disastrous consequences. The obvious examples in this regard are the investment banking firms (e.g., Lehman Brothers) in which relentless focus on short-term business results and bonuses resulted in corruption permeating the entire organization and eventually resulted in their demise.

Secondly, the hierarchical level at which the *bad apples* are located has significant impact on the corruption spreading dynamics. As one would expect, the higher the level at which the *bad apples* are located, the faster and wider the spread of corruption in the organization. However, if the organization is able to hire employees who are less likely to succumb to corrupt influences, then the differential impact of hierarchical level is even more pronounced, i.e., junior-level *bad apples* have a much lower impact on corruption spreading dynamics than senior-level *bad apples*. This implies that the testing of senior-level job applicants with regard to ethics should be conducted more rigorously than for junior-level job applicants. However, once again it seems that in real-life the opposite is true, and junior-level positions (e.g., sales staff in retail organizations, or tellers in consumer banking) are subjected to greater scrutiny on ethics than senior-level positions (e.g., sales managers, bank branch managers). Also, the socialization processes for junior employees (e.g., management trainees) is usually far more formal, rigorous and comprehensive than for senior employees. Therefore, if a senior employee is carrying a "corruption virus" with him or her, the lack of a rigorous socialization process will allow the virus to be retained and it could initiate a corruption contagion in the future.

Thirdly, the presence of potential whistle-blowers is an important antidote to corruption spreading. Even if 5% of the workforce are potential whistle-blowers then the chances of the corruption being inhibited are very high, and if this number can be raised to around 25%, then the impact of *bad apples* on corruption spreading will be negligible. Thus, our simulations provide some indication of what would be a reasonable target to achieve in terms of fostering whistle-blowers in an organization. The lower bound, i.e., 5% is not an impossibly tall order for an organization to shoot for and if organizations create a climate conducive to whistle-blowing then at least the lower bound could be achieved.

# IV. DISCUSSION AND CONCLUSIONS

Our paper bridges two important research streams, organizational corruption and social contagion. With regard to organizational corruption research, we complement the rich existing conceptual work by formally modeling corruption contagion. We have quantitatively depicted the time-evolution of corruption across organizations with different organization structures and estimated the critical thresholds at which the corruption would irrevocably contaminate the organization and the proportion of whistle-blowers needed to inhibit the spread of corruption. As mentioned earlier, this is one of the few attempts to study the role of organization structure in influencing ethical actions. Our paper also contributes to the literature on social contagion, by modeling intra-organizational contagion across individuals, an area that has been relatively neglected. Further, we are not aware of other social contagion studies that have included a parameter that impedes the contagion, such as the potential whistle-blowers in our study.

One of the benefits of formal modeling is unanticipated implications (Adner et al. 2009), and we plan to further investigate some of the intriguing results in this paper, in particular, the impact of taller versus flatter structures on corruption spreading dynamics. Also, in organizations there are power differentials among individuals in a dyad, and power asymmetry would influence the corruption probability. This means that it would be easier for a superior to influence a subordinate to adopt a corrupt practice, than it would be for a subordinate to influence a superior, or even a peer. In terms of our simulation studies, this implies that rather than have a fixed corruption probability we could have a range of corruption probabilities which reflect power asymmetries.

Our paper is generative in terms of future research. For instance, now that we have a ball-park estimate of the critical threshold of "corruption probability" of individuals, we could conduct studies using threshold models rather than Poisson models. The base-rate "corruption probability" of individuals itself could be estimated and triangulated through laboratory experiments. This lab base rate could then be adjusted to account for the effect of factors like the organization's climate for ethics (Dickson, Smith, Grojean, & Ehrhart, 2001) or ethical work climate (Victor & Cullen 1988).

In this paper, we have varied the corruption probability and considered the evolution of the corruption contagion emanating from a single *bad apple*. In future work, we could keep the corruption probability fixed at a low value, and vary the number of *bad apples*, who could be randomly distributed through the organization to investigate the contagion dynamics. This approach would reflect the situation wherein although the majority of the employees are high in integrity, there are some *bad apples* who should have been rejected but have been hired.

Considering the interesting results with regard to organization structure, i.e., that the compartmentalization emanating from flatter structures results in slower spreading and lower overall contamination, we could model contagion across other underlying network structures. For instance, core-periphery structures, which are characterized by a dense, cohesive core and a sparse, unconnected periphery (Abrahamson & Rosenkopf 1997; Borgatti & Everrett 1999), are not only considered representative of organization structure (where the top management team is the core, and the rest of the organization is the periphery) but also rife in structural holes, and modeling corruption contagion over this network structure might prove insightful.

We could also consider the impact of power asymmetries and other situational factors on the effectiveness of whistle-blowing. Combining the results of our simulations, since corruption contagion is more inimical at senior levels, and presence of a small number of whistle-blowers is an effective antidote, we could explore the impact of varying the hierarchical level location of the whistle-blowers in future research, instead of simply taking a random sample as we have done here.